\documentclass[11pt]{article}
\usepackage{enumitem}
 \usepackage{amsfonts}
\usepackage{amsmath}
\usepackage{graphicx}
\usepackage{fancyhdr}
\usepackage{amsthm}
\usepackage{txfonts}
\usepackage{amssymb}
\usepackage{listings}
\usepackage{amscd}
\usepackage{float}
\usepackage{color}
\usepackage{authblk}
\newtheorem{definition}{Definition}[section]

\newtheorem{lemma}{Lemma}[section]
\newtheorem{proposition}{Proposition}[section]
\newtheorem{remark}{Remark}[section]

\begin{document}

\title{{Continuous time mean-variance-utility portfolio problem and its equilibrium strategy}}
\author{{Ben-Zhang Yang$^a$, Xin-Jiang He$^b$\footnote{Corresponding author. E-mail address: xinjiang@uow.edu.au.}, and Song-Ping Zhu$^b$}\\
{\small\it a. Department of Mathematics, Sichuan University, Chengdu, Sichuan 610064, P.R. China}\\

{\small\it b. School of Mathematics and Applied Statistics, University of Wollongong NSW 2522, Australia}\\
}
\date{}
\maketitle
\vspace*{-9mm}
\begin{center}
\begin{minipage}{5.5in}
{\bf Abstract.}
In this paper, we propose a new class of optimization problems, which maximize the terminal wealth and accumulated consumption utility subject to a mean variance criterion controlling the final risk of the portfolio.  The multiple-objective optimization problem is firstly transformed into a single-objective one by introducing the concept of overall ``happiness" of an investor defined as the aggregation of the terminal wealth under the mean-variance criterion and the expected accumulated utility, and then solved under a game theoretic framework. We have managed to maintain analytical tractability; the closed-form solutions found for a set of special utility functions enable us to discuss some interesting optimal investment strategies that have not been revealed before in literature.
\\ \ \\

\end{minipage}
\end{center}
\section{\textcolor{blue}{Introduction}}

\textcolor{blue}{The} optimal portfolio selection problem is essentially to achieve a balance between uncertain returns and risks, for which the mean-variance methodology has become one of the most important tools, ever since Markowitz's pioneering work \cite{Markowitz52} on a static investment model. This approach conveys a nice and elegant idea, maximizing the expected return at a given level of risk or minimizing the risk at a given level of expected return, and has been applied to finance practice (see \cite{He17,Ma191,Ma192,Pun18,Shen15,Yan19,Yang19,Yang20,Yang20a,Zhu11}).

Although Markowitz's work is theoretically very appealing, it only provides results based on a single-period (static) model, in which the investors can only make a decision at the very beginning, while they are not allowed to make any adjustments before the investment period ends. In fact, dynamic mean-variance optimization is not a trivial task, as the Bellman's dynamic programming principle can not be directly applied to this kind of path-dependent optimization problems due to the nonlinearity of the variance operator. In this sense, such kind of problems is usually referred to as time-inconsistent problems and the corresponding solutions are called time-inconsistent solutions.

A natural approach to solve time-inconsistent problems is to optimize the objective with a fixed initial point under the mean-variance criterion, and the derived solution is regarded as an optimal pre-commitment solution. For instance, with the market being complete and continuous timewise, various results have been presented for the variance-minimizing policy using martingale methods, given that the expected terminal wealth is equal to a certain level (see \cite{Bielecki05,Cvitanic08,Cvitanic04,Zhao02}). In contrast, Cochrane \cite{Cochrane08} derived the optimal investment policy that minimizes the ``long-term" variance of portfolio returns subject to the constraint that the long-term mean of portfolio returns equals to a pre-specified target level under an incomplete market. This approach has also been applied in futures trading strategies by Duffie and Richardson \cite{Duffie91} through setting a mean-variance objective at the initial date. They obtained a pre-commitment solution , which also solves the optimal problem with a quadratic objective for some specific parameters. A similar approach developed for continuous time complete-market settings has also been widely discussed (see \cite{Leippold04,Li00,Lim02,Zhou00}).

However, Basak and Chabakauri \cite{Basak10} challenged the pre-commitment assumption \cite{Zhou00}, and assumed investors are sophisticated in the sense that they will maximize their mean-variance objective over time considering all future updates, instead of finding an optimal solution at a fixed given time moment. Following this, Kryger and Steffensen \cite{Kryger10} worked under the Black-Scholes framework without the pre-commitment assumption, and showed that the optimal strategy derived for a mean-standard deviation investor is to take no risk at all. \textcolor{blue}{The latest contribution to the relevant literature was made as presented in \cite{Bjork12}, where mean-variance optimization problems were considered under a game theoretic framework \cite{Bjork09}, and the optimal strategies were derived in the context of sub-game perfect Nash equilibrium.}

Although all the work mentioned above is very appealing, consumption was usually not considered in mean-variance optimization problems, which is not consistent with what is happening in practice, as the decisions made for consumption would naturally have an impact on the optimality of the investment strategy. Thus, Christiansen \& Steffensen \cite{Christiansen13} and Kronborg \& Steffensen \cite{Kronborg15} started to incorporate consumption choices into the mean-variance problem, investigating the optimal investment-consumption problem together with the mean-variance criterion. Unfortunately, the optimal consumption strategies derived under their particular model assumptions and the formulations of the problem have led to a rather absurd conclusion that an investor could suddenly be required to switch his/her consumption strategy from consuming as much as possible to as little as possible, in order to achieve the ``optimal" objective set at the beginning of the investment period. It is this fundamental flaw in the state-of-the-art model frame of the portfolio selection problem that has prompted us to propose a new model frame that would not only eliminate this rather strange behavior but also maintain mathematical tractability so that a more rational investment behavior can be discussed economically for some simple utility functions through analytical closed-form solutions.

A new class of continuous-time portfolio selection problems is proposed in this paper, which combine maximizing the terminal wealth under the mean-variance criterion and maximizing accumulated consumption utility together. \textcolor{blue}{Different from previously presented literature \cite{Christiansen13, Kronborg15}, they added the consumption term directly back to the terminal wealth in the objective value function of the optimization problem under mean-variance criterion, while we introduce the concept of overall "happiness" of an investor, which is measured by the aggregation of the terminal wealth under the mean-variance criterion and the expected accumulated utility, using a consumption preference parameter.} Amazingly, the newly formulated optimization problem preserves the analytical tractability under a continuous-time game theoretic framework, \textcolor{blue}{and the analytical optimal continuous investment and consumption strategies derived in the sense of equilibrium admit intuitive economic explanation.}

The rest of this paper is organized as follows. Section 2 reviews the classical mean-variance problem and proposes the new portfolio selection problem. In Section 3, we analytically derive the optimal strategies based on the definition of the equilibrium strategy. Explicit solutions to the optimal strategies are then presented in Section 4. Numerical examples and discussions are provided in Section 5, followed by some concluding remarks given in the last section.

\section{The portfolio selection problem}

\textcolor{blue}{In this section, we first briefly review the classical continuous-time mean-variance portfolio selection model, and then propose our new model by introducing consumption into the original model.}

\subsection{The classical mean-variance portfolio problem}

We now assume that we work under the standard Black-Scholes market, where an investor has access to a \textcolor{blue}{risk-free bank account $M$ and a stock $S$} whose dynamics can be specified as
\begin{equation}
\begin{aligned}
dM(t)&=rM(t)dt, &&\quad M(0)=1,\\
dS(t)&=\mu S(t)dt+\sigma S(t) dB(t), && \quad S(0)=s_0>0.
\end{aligned}
\end{equation}
Here, $r>0$, $\mu$ and $\sigma$ are constants, and it is assumed that $\mu>r$. The process $B(t)$ is a standard Brownian motion on the probability space $(\Omega, \mathcal{F},\mathbb{P})$ with the filtration $\sigma\{B(s);0\leq s \leq t\}$, $\forall t \in [0,T]$.

We also assume that the investor in this market needs to make investment decisions on a finite time horizon $[0,T]$, and he/she \textcolor{blue}{allocates a proportion $\pi(t)$ and $1-\pi(t)$ of his wealth into the stock and bank account}, respectively, at time $t$. Let $X^{\pi}(t)$ be the wealth of the investor at time $t$ following the investment strategy $\pi(\cdot)$ with an initial wealth of $x_0$ at time $0$. \textcolor{blue}{In other words, it is required that $X^{\pi}(t)=\pi(t)S(t)+(1-\pi(t))M(t)$. Therefore, the dynamic of the investor's wealth follows}
\begin{equation}\label{wealth2}
\left\{
\begin{aligned}
dX^{\pi}(t)&=[(r+\pi(t)(\mu-r))X^{\pi}(t)]dt+\pi(t)\sigma X^{\pi}(t)dB(t), \quad t \in [0,T),\\
X(0)&=x_0>0.
\end{aligned}
\right.
\end{equation}

If $L^2_{\mathcal{F}}(0,T;R)$ denotes the set of all $R$-valued, measurable stochastic process $f(t)$ adapted to $\{F_t\}_{t\geq 0}$ such that $E\left[\int_{0}^{T}f^2(t)dt\right]<\infty$, the classical continuous time mean-variance portfolio optimization problem is stated below.

\begin{definition}(\cite{Zhou00})
A portfolio strategy $\pi(\cdot)$ is admissible if $\pi(\cdot) \in L^2_{\mathcal{F}}(0,T;R)$.
\end{definition}

\begin{definition}(\cite{Zhou00})
The continuous time mean-variance portfolio optimization problem is a multi-objective optimization problem, which is defined as
\begin{equation}\label{max0}
\begin{aligned}
&\min_{\pi(\cdot)} &&(V_1(\pi(\cdot)),V_2(\pi(\cdot)))\textcolor{blue}{\equiv} (-E(X(T)),Var(X(T))),\\
&s.t. && \left\{
\begin{aligned}
&\pi(\cdot) \in L^2_{\mathcal{F}}(0,T;R),\\
&(X(\cdot),\pi(\cdot)) ~satisfy ~Equation ~ \eqref{wealth2}.
\end{aligned}
\right.
\end{aligned}
\end{equation}
\end{definition}

The optimization problem \eqref{max0} can be transformed into a single-objective optimization problem by introducing a weight parameter $\gamma$ such that the new objective becomes the weighted average of two original objectives using mild convexity conditions \cite{Zeleny}
\begin{equation}\label{max1}
\begin{aligned}
&\min_{\pi(\cdot)} &&V_1(\pi(\cdot))+\frac{\gamma}{2} V_2(\pi(\cdot)) \equiv -E(X(T))+\frac{\gamma}{2}Var(X(T)),\\
&s.t. && \left\{
\begin{aligned}
&\pi(\cdot) \in L^2_{\mathcal{F}}(0,T;R),\\
&(X(\cdot),\pi(\cdot)) ~satisfy ~Equation ~ \eqref{wealth2},
\end{aligned}
\right.
\end{aligned}
\end{equation}
where the weight parameter satisfies $\frac{\gamma}{2}>0$. In fact, this particular optimization problem \eqref{max1} has been extensively studied in the past 20 years with various theoretical results, numerical algorithms, and applications being available in the literature. Interested readers are referred to \cite{Basak10,Zhao02,Zhou00} and the references therein for more details.

\subsection{The mean-variance-utility consumption and investment problem}

It should be pointed out that the classical mean-variance optimization problem (\ref{max0}) or the transformed optimization problem (\ref{max1}) \textcolor{blue}{does not take into account the investor's income and consumption. The main possible reason is that the incorporation of the consumption choices in the classical mean-variance problem could destroy the tractability of the original problem.} However, this is apparently not appropriate as it is not consistent with real situations, and thus we \textcolor{blue}{consider a natural extension of the mean-variance problem.} Assume that the investor possesses a continuous deterministic income rate $l(t)$, and chooses a non-negative consumption rate $c(t)$. Under these assumptions, the dynamic of the investor's wealth can be derived as
\begin{equation}\label{wealth}
\left\{
\begin{aligned}
dX^{c,\pi}(t)&=[(r+\pi(t)(\mu-r))X^{c,\pi}(t)+l(t)-c(t)]dt+\pi(t)\sigma X^{c,\pi}(t)dB(t), \quad t \in [0,T),\\
X(0)&=x_0>0.
\end{aligned}
\right.
\end{equation}

Obviously, after incorporating the consumption into the mean-variance problem, the investor is also seeking for his/her maximum utility through the consumption choices, apart from the mean-variance type objective as specified in the optimization problem \eqref{max1}\footnote{Although the two wealth dynamics \eqref{wealth2} and \eqref{wealth} are different, the optimal solution to the mean-variance problem \eqref{max1} with \eqref{wealth2} and that to the mean-variance problem \eqref{max1} with \eqref{wealth} are the same. In particular, by setting $\rho=0$ in Proposition 3.1 of \cite{Kronborg15}, one can easily guarantee the optimal strategy, and the results in \cite{Kronborg15} also show that the deterministic cash flow does not bring any possible adjustment to the investment strategy under the mean-variance criterion.}. In other words, the investor again faces a dual-objective optimization problem; he/she wants to achieve the maximum accumulated utility over a choice of consumption, while at the same time minimizing investment risk by considering a mean-variance objective over terminal wealth $X(T)$. In this case, we need a measurement for the utility obtained through consumption. With $\rho$ representing a constant discounting rate and $U(\cdot)$ denoting an utility function, the accumulated utility of the investor through his/her continuous consumption on time interval $[t,T]$ can be defined as
\begin{equation}
V_3^{c,\pi}(t,x)=E\left[\int_{t}^{T}e^{-\rho(s-t)}U(c(s))ds \right],
\end{equation}
where $E(\cdot)$ denotes taking the expectation \textcolor{blue}{and utility function $U(\cdot)$ is continuous, concave and increasing with $U(0)=0$.} It is this new optimal portfolio selection problem, named as ``mean-variance-utility consumption and investment optimization problem", that is presented below.
\begin{definition}
A new mean-variance-utility consumption and investment optimization problem can be formulated as
\begin{equation}\label{maxu}
\begin{aligned}
&\max &&\left[V_1^{c,\pi}(t,x),V_2^{c,\pi}(t,x),V_3^{c,\pi}(t,x)\right]\equiv \left[ E(X(T)),-Var(X(T)),E\left(\int_{t}^{T}e^{-\rho(s-t)}U(c(s))ds\right)\right]\\
&s.t. && \left\{
\begin{aligned}
&c(\cdot),\pi(\cdot) \in L^2_{\mathcal{F}}(0,T;R),\\
&(\textcolor{blue}{X(\cdot)},c(\cdot),\pi(\cdot))~ satisfy~ Equation~ \eqref{wealth}.
\end{aligned}
\right.
\end{aligned}
\end{equation}
\end{definition}

Similarly to what have been presented in the previous subsection, the optimization problem \eqref{maxu} can also be converted into a single-objective optimization problem
\begin{equation}\label{maxu2}
\begin{aligned}
&\max_{c(\cdot),\pi(\cdot)} && E(X(T))-\frac{\gamma}{2}Var(X(T))+\beta E\left(\int_{t}^{T}e^{-\rho(s-t)}U(c(s))ds\right)\\
&s.t. && \left\{
\begin{aligned}
&c(\cdot), \pi(\cdot) \in L^2_{\mathcal{F}}(0,T;R),\\
&(X(\cdot),c(\cdot),\pi(\cdot)) ~satisfy ~Equation ~ \eqref{wealth},
\end{aligned}
\right.
\end{aligned}
\end{equation}
where $\beta$ is a positive constant. It should be noted that the problem \eqref{maxu2}  degenerates to the classical mean-variance portfolio selection model \eqref{max1} when $\beta$ approaches zero.

Clearly, the parameter $\beta$ can be treated as a trade-off between acquiring more terminal wealth in the mean-variance sense and achieving more accumulated utility through consumption; the larger the value of $\beta$ is, the more inclined the investor is to consume to maximize his/her accumulated utility. In this sense, $\beta$ is actually a consumption preference parameter. It should also be noted that the introduced parameter $\beta$ can be regarded as a conversion operator that converts the utility units to wealth units.

It should be particular emphasized that although Christiansen and Steffensen \cite{Christiansen13}; Kronborg and Steffensen \cite{Kronborg15} have already tried to incorporate the consumption into the mean-variance framework, the problem they discussed is essentially different from our new problem \eqref{maxu2}. This is because they have directly added the accumulated consumption to the terminal wealth to formulate an ``adjusted" terminal wealth and considered the mean-variance optimization of the adjusted wealth, while we have distinguishably introduced parameter $\beta$ so that the mean-variance of the terminal wealth and the accumulated consumption utility can be added together to form a new objective. In this way, the new objective can be economically interpreted as the overall ``happiness" of an investor towards his/her investment return as well as the undertaken risk level during the time period $[0,T]$.

The investor aims at achieving the maximum overall happiness through the combination of maximizing the terminal wealth with the mean-variance criteria and the consumption utility, leading to a new class of mean-variance-utility optimization problems. What our new model suggests is that an investor should not proportionally consume as suggested by the Merton's classic framework and she/he should also consider the balance of the total wealth management under the Markowitz's mean-variance criterion. The fundamental reason is because now he/she still wants to minimize his/her total investment risk at the end of the investment period, while maximizing his/her expected return and accumulated consumption utility. Of course, it is interesting to explore whether the introduction of consumption utility would affect the previous structure of the investment strategy reported in \cite{Basak10,Bjork12,Kronborg15}. The new challenge here is to solve the new optimal portfolio selection problem \eqref{maxu2}, which will be discussed in the next section.

\section{Optimal portfolio selection strategy}

\textcolor{blue}{Having successfully established a new optimal portfolio selection problem in \eqref{maxu2}, a natural question is whether or not there exists a solution to the optimal portfolio selection strategy, and how it can be derived if it does exist. This is a challenging problem because we are not able to find time-consistent solutions in the sense that the condition for the Bellman Optimality Principle no longer holds, given that the law of iterated expectations does not apply for a given strategy. In this section, we attempt to seek an optimal solution to problem \eqref{maxu2} in the sense of time inconsistency.}

\textcolor{blue}{Similar to \cite{Bjork09,Bjork12}, a natural way for an investor to deal with any time inconsistent problem is to solve the problem by setting $t=0$, and the investor will follow the resulting optimal strategy during the finite time horizon. This is the so-called pre-commitment control, i.e., the investor pre-commits at a fixed time moment. However, one of the main drawbacks for the optimal solution with pre-commitment is that it will not be optimal for the control problem at any time $t>0$.}

In fact, most investors in practice would assign same weights to all time instances, implying that they are looking for an optimal strategy that is optimal from the point of view at any time $t$ during the considered time horizon instead of time $0$. Therefore, instead of seeking  a pre-commitment solution, our problem is considered under a game theoretic framework without pre-commitment, which was introduced in \cite{Bjork09,Bjork12} and developed by Kronborg and Steffensen \cite{Kronborg15} as well as Kryger et al. \cite{Kryger19}.

To solve the optimization problem \eqref{maxu2}, we consider a more general optimization problem with a discount factor as follows:
\begin{equation}\label{maxu22}
\begin{aligned}
&\max_{c(\cdot),\pi(\cdot)} && E(e^{-\delta(T-t)}X(T))-\frac{\gamma}{2}Var(e^{-\delta(T-t)}X(T))+\beta E\left(\int_{t}^{T}e^{-\rho(s-t)}U(c(s))ds\right)\\
&s.t. && \left\{
\begin{aligned}
&c(\cdot), \pi(\cdot) \in L^2_{\mathcal{F}}(0,T;R),\\
&(X(\cdot),c(\cdot),\pi(\cdot)) ~satisfy ~Equation ~ \eqref{wealth},
\end{aligned}
\right.
\end{aligned}
\end{equation}
where $\delta$ is a discount rate, \textcolor{blue}{$\gamma$ is the risk-aversion coefficient, $\beta$ is a positive trade-off parameter. } Obviously, the optimization problem \eqref{maxu22} degenerates to the original one \eqref{maxu2} when the investor requires a discount rate of 0.

The equilibrium strategy under the continuous-time game theoretic equilibrium for the problem \eqref{maxu22} can be defined below.
\begin{definition}\label{def1}
Consider a strategy $(c^*,\pi^*)$ and a fixed point $(c,\pi)$. For a fixed number $h>0$ and an initial point $(t,x)$, we define the strategy $(\widetilde{c}_h,\widetilde{\pi}_h)$ as
\begin{equation}\label{pi0}
(\widetilde{c}_h(s),\widetilde{\pi}_h(s))=\left\{
\begin{aligned}
&(c,\pi), && \text{for} \quad  t\leq s< t+h,\\
&(c^*(s),\pi^*(s)), &&\text{for} \quad t+h \leq s <T.
\end{aligned}
\right.
\end{equation}
If
\begin{equation}
\lim_{h\rightarrow 0} \inf \frac{1}{h}\left(f^{c^*,\pi^*}(t,x,y^{c^*,\pi^*},z^{c^*,\pi^*},w^{c^*,\pi^*})-f^{\widetilde{c}_h,\widetilde{\pi}_h}(t,x,y^{\widetilde{c}_h,\widetilde{\pi}_h},z^{\widetilde{c}_h,\widetilde{\pi}_h},w^{\widetilde{c}_h,\widetilde{\pi}_h})\right)\geq 0
\end{equation}
for all $(c,\pi) \in \mathbb{R}_+ \times \mathbb{R}$, where $f$ is  \textcolor{blue}{a general object function} and
\begin{equation}\label{yzw0}
\begin{aligned}
&y^{c,\pi}:=y^{c,\pi}(t,x)=E\left[\left.e^{-\delta(T-t)}X^{c,\pi}(T)\right|X(t)=x\right],\\
&z^{c,\pi}:=z^{c,\pi}(t,x)=E\left[\left.\left(e^{-\delta(T-t)}X^{c,\pi}(T)\right)^2\right|X(t)=x\right],\\
&w^{c,\pi}:=w^{c,\pi}(t,x)=E\left[\left.\int_{t}^{T}e^{-\rho(T-t)}U(c(s))\right|X(t)=x\right],\\
\end{aligned}
\end{equation}
then $(c^*,\pi^*)$ is \textcolor{blue}{called} an equilibrium strategy.
\end{definition}

The equilibrium strategy \textcolor{blue}{under Nash Equilibrium Criteria} defined in Definition \ref{def1} is a time-inconsistent solution to the control problem \eqref{maxu2}, which is essentially different from time-consistent solutions discussed in the context of optimization \cite{Basak10}. If we denote $(c^*,\pi^*)$ as the equilibrium strategy satisfying Definition \ref{def1}, and let $V$ be the  the corresponding value function with the equilibrium strategy, we can obtain
\begin{equation}\label{op1}
V(t,x)=f^{c,\pi}(t,x,y^{c^*,\pi^*},z^{c^*,\pi^*},w^{c^*,\pi^*}).
\end{equation}
Clearly, our problem is to search for the corresponding optimal strategies and the optimal value function $f:[0,T]\times \mathbb{R}^4\rightarrow R$ as a $\mathcal{C}^{1,2,2,2,2}$ function of the form
\begin{equation}\label{ff}
f^{c^*,\pi^*}(t,x,y^{c,\pi},z^{c,\pi},w^{c,\pi})=y-\frac{\gamma}{2}(z-y^2)+\beta w, \quad (c,\pi)\in \mathcal{A},
\end{equation}
where $\mathcal{A}$ is the class of admissible strategies to be defined below.
As pointed in \cite{Kryger19}, the investor continuously deviates from this strategy and thus does not actually achieve any of the determined supremums. \textcolor{blue}{Instead, the investor concentrates on determining the equilibrium control law, as introduced in \cite{Bjork09,Bjork12} and \cite{Kryger10}. The desired investment strategy is determined so that it maximizes the present objective at any time moment $t$, under the restriction that the future strategy is assumed to be given.} In other words, the strategy is determined through backward recursion, and thus this recursively optimal solution under equilibrium control law is also regarded as the optimal control (see \cite{Kryger10,Kryger19}).

Before we are able to present the optimal solution, some preliminaries need to be outlined. In particular, we establish an extension of the HJB equation \textcolor{blue}{which will be discussed in detail later in Lemma \ref{lemma2}} for the characterization of the optimal value function and the corresponding optimal strategy, so that the stochastic problem can be transformed into a system of deterministic differential equations and a deterministic point-wise minimization problem.

\textcolor{blue}{Let $\mathcal{A}$  be the set of admissible strategies that contains all strategies $(c,\pi)$ satisfying the following two conditions: i) there exist solutions to the partial differential equations \eqref{Y1}-\eqref{W1}; ii) the stochastic integral in \eqref{Y2} ,\eqref{Z3}, \eqref{W3} and \eqref{F5} are martingales. In the same way as in \cite{Kronborg15}, the equilibrium strategy we solve below is restricted in set $\mathcal{A}$.} Then, we can prove the following two lemmas.
\begin{lemma}\label{lemma1}
\textcolor{blue}{If there exist three functions $Y=Y(t,x)$, $Z=Z(t,x)$ and $W=W(t,x)$ such that
\begin{equation}\label{Y1}
\left\{
\begin{aligned}
Y_t(t,x)&=-[(r+\pi(\mu-r))x+l-c]Y_x(t,x)-\frac{1}{2}\pi^2\sigma^2x^2Y_{xx}(t,x)+\delta Y(t,x),\\
Y(T,x)&=x,
\end{aligned}
\right.
\end{equation}
\begin{equation}\label{Z1}
\left\{
\begin{aligned}
Z_t(t,x)&=-[(r+\pi(\mu-r))x+l-c]Z_x(t,x)-\frac{1}{2}\pi^2\sigma^2x^2Z_{xx}(t,x)+2\delta Z(t,x),\\
Z(T,x)&=x^2,
\end{aligned}
\right.
\end{equation}
and
\begin{equation}\label{W1}
\left\{
\begin{aligned}
W_t(t,x)&=-[(r+\pi(\mu-r))x+l-c]W_x(t,x)-\frac{1}{2}\pi^2\sigma^2x^2W_{xx}(t,x)-e^{-\rho t}U(c),\\
W(T,x)&=0,
\end{aligned}
\right.
\end{equation}
where $(c,\pi)$ is an arbitrary admissible strategy, then
\begin{equation}\label{s1}
Y(t,x)=y^{c,\pi}(t,x),\quad Z(t,x)=z^{c,\pi}(t,x), \quad W(t,x)=w^{c,\pi}(t,x),
\end{equation}
where $y^{c,\pi}$, $z^{c,\pi}$  and $w^{c,\pi}$ are given by  \eqref{yzw0}.}
\end{lemma}
\begin{proof}
\textcolor{blue}{See Appendix A for the proof.}
\end{proof}

\begin{lemma}\label{lemma2}
If there exists a function $F=F(t,x)$ such that
\begin{equation}\label{F1}
\left\{
\begin{aligned}
&F_t=\inf_{c,\pi \in \mathcal{A}} \left\{-[(r+\pi(\mu-r))x+l-c](F_x-Q)-\frac{1}{2}\pi^2\sigma^2x^2(F_{xx}-K)+J\right\},\\
&F(T,x)=f^{c,\pi}(T,x,x,x^2,0),
\end{aligned}
\right.
\end{equation}
where $Q=f_x^{c^*,\pi^*}$,
\begin{equation}\label{e30}
\begin{aligned}
K=&f_{xx}^{c^*,\pi^*}+f_{yy}^{c^*,\pi^*}(F^{(1)}_{x})^2++f_{zz}^{c^*,\pi^*}(F^{(2)}_{x})^2+f_{ww}^{c^*,\pi^*}(F^{(3)})^2+2f_{xy}^{c^*,\pi^*}F^{(1)}_x+2f_{xz}^{c^*,\pi^*}F^{(2)}_x\\
&+2f_{xw}^{c^*,\pi^*}F^{(3)}_x+2f_{yz}^{c^*,\pi^*}F^{(1)}_xF^{(2)}_x+2f_{yw}^{c^*,\pi^*}F^{(1)}_xF^{(3)}_x+2f_{zw}^{c^*,\pi^*}F^{(2)}_xF^{(3)}_x
\end{aligned}
\end{equation}
and
\begin{equation}\label{e31}
J=f_t^{c^*,\pi^*}+f_y^{c^*,\pi^*}\delta F^{(1)}+2f_z^{c^*,\pi^*}\delta F^{(2)}-f_w^{c^*,\pi^*}e^{-\rho t}U(c(t)).
\end{equation}
with
$$F^{(1)}=y^{c^*,\pi^*}(t,x), \quad F^{(2)}=z^{c^*,\pi^*}(t,x), \quad F^{(3)}=w^{c^*,\pi^*}(t,x),$$
then
$$F(t,x)=V(t,x),$$
where $V$ is the optimal value function defined by \eqref{op1}.
\end{lemma}
\begin{proof}
\textcolor{blue}{See Appendix B for the proof.}
\end{proof}

\begin{remark}
The representation corresponds to the pseudo-Bellman equation \eqref{F1}, originally presented in \cite{Kryger10} and applied in Theorem 2.1 of \cite{Kronborg15}, calls for an optimization across strategies, whereas the whole point of dynamic programming is to appeal only to optimization across vectors. Therefore, the optimal solution solved by this approach belongs to the subspace of $\mathcal{A}$. Similar to Bj\"{o}rk and Murgoci \cite{Bjork09}, the optimality of the our obtained strategy can also be confirmed in the sense of equilibrium.
\end{remark}

\section{Optimal consumption and investment}

\textcolor{blue}{In this section, we present the optimal solutions to the optimal portfolio selection problem \eqref{maxu2} based on the results derived in the previous section, and some detailed discussions are provided to illustrate the behaviour of the optimal strategies.}

A candidate strategy for the optimal value function \eqref{F1} can be derived by simply differentiating \eqref{F1} with respect to $c$ and $\pi$, respectively. \textcolor{blue}{Fortunately, we notice that the terms containing $c$ and $\pi$ are independent of each other.} This leads to
\begin{equation}
\frac {\partial}{\partial c}\left(c(F_x-Q)-f_we^{-\rho t}U'(c)\right)=0
\end{equation}
and
\begin{equation}
\frac {\partial}{\partial \pi}\left(-\pi(\mu-r)x(F_x-Q)-\frac{1}{2}\pi^2\sigma^2x^2(F_{xx}-K)\right)=0.
\end{equation}
A further simplification then yields
\begin{equation}\label{o1}
\left\{
\begin{aligned}
c^*&=[U']^{-1}\left(\frac{F_x-Q}{f_w}e^{-\rho t}\right),\\
\pi^*&=-\frac{\beta-r}{x\sigma^2}\frac{F_x-Q}{F_{xx}-K},
\end{aligned}
\right.
\end{equation}
where $[f]^{-1}(\cdot)$ is the inverse function of $f$, and stars denote that they are the optimal strategies.

Substituting the corresponding objective form
\begin{equation}\label{ff2}
f(t,x,y,z,w)=y-\frac{\gamma}{2}(z-y^2)+\beta w
\end{equation}
into \eqref{e30} and \eqref{e31} gives
\begin{equation}\label{z}
Q=0, \quad K=\gamma(F_x^{(1)})^2, \quad J=\delta F^{(1)}-\gamma\delta\left(F^{(2)}-(F^{(1)})^2\right)-\beta e^{-\rho t}U(c).
\end{equation}

\textcolor{blue}{Given the linear structure of the dynamics \eqref{Y1}, \eqref{Z1} and \eqref{W1}, as well as of the boundary conditions, $F$, $F^{(1)}$ and $F^{(3)}$ can be written in the following form:}
\begin{equation}\label{aa1}
F(t,x)=A(t)x+B(t), \quad
F^{(1)}(t,x)=a(t)x+b(t), \quad
F^{(3)}(t,x)=p(t)x+q(t),
\end{equation}
which then naturally leads to $F^{2}$ being written in the form
\begin{equation}\label{aa2}
F^{(2)}(t,x)=\frac{2}{\gamma}\left[a(t)x+b(t)+\beta\left[p(t)x+q(t)\right]-\left[A(t)x+B(t)\right]\right]+\left[a(t)x+b(t)\right]^2.
\end{equation}
\textcolor{blue}{Similar method and the general verification results can be found in \cite{Bjork09,Bjork12, Kronborg15}.}
Now, the substitution of \eqref{aa1} and \eqref{aa2} into \eqref{z} results in
\begin{equation}\label{z2}
Q=0, \quad K=\gamma\left(a(t)\right)^2,
\end{equation}
and
\begin{equation}\label{J}
J=\delta\left[a(t)x+b(t)\right]-2\delta\left[a(t)x+b(t)+\beta\left[p(t)x+q(t)\right]-\left[A(t)x+B(t)\right]\right]-\beta e^{-\rho t}U(c(t)),
\end{equation}
with which the optimal strategy \eqref{o1} becomes
\begin{equation}\label{o2}
\left\{
\begin{aligned}
c^*&=[U']^{-1}\left(\beta^{-1}e^{-\rho t}A(t)\right),\\
\pi^*&=\frac{1}{\gamma}\frac{\mu-r}{x\sigma^2}\frac{A(t)}{a^2(t)}.
\end{aligned}
\right.
\end{equation}
If we further substitute \eqref{z2}, \eqref{J} and \eqref{o2} into \eqref{Y1} - \eqref{W1} and \eqref{F1} with the corresponding terminal conditions, it is straightforward to show that
\begin{equation}
\left\{
\begin{aligned}
A_tx+B_t=&-rxA-\frac{1}{2\gamma}\frac{(\mu-r)^2A^2}{\sigma^2a^2}-lA+c^*A-\beta e^{-\rho t}U(c^*)+\delta(ax+b)\\
         &-2\delta(ax+b-Ax-B+\beta(px+q)),\\
a_tx+b_t=&-rxa-\frac{1}{\gamma}\frac{(\mu-r)^2A}{\sigma^2a}-la+c^*a+\delta(ax+b), \\
p_tx+q_t=&-rxp-\frac{1}{\gamma}\frac{(\mu-r)^2Ap}{\sigma^2a^2}-lp+c^*p- e^{-\rho t}U(c^*(t)),
\end{aligned}
\right.
\end{equation}
with terminal conditions $A(T)=a(T)=1$ and $B(T)=b(T)=p(T)=q(T)=0$.

After some further simplifications, we can obtain
$
A(t)=a(t)=e^{(r-\delta)(T-t)},
$
$p(t)=0$, $q(t)=\int_{t}^{T}e^{-\rho s}U(c^*(s))ds$,
and
\begin{equation}\label{BB}
\begin{aligned}
b(t)&=e^{\delta t}\int_{t}^{T}\left[\frac{1}{\gamma}\frac{(\mu-r)^2}{\sigma^2}+l(s)e^{(r-\delta)(T-s)}-c^*(s)e^{(r-\delta)(T-s)}\right]e^{-\delta s}ds,\\
B(t)&=e^{2\delta t}\int_{t}^{T}\left[\frac{1}{2\gamma}\frac{(\mu-r)^2}{\sigma^2}+l(s)e^{(r-\delta)(T-s)}-c^*(s)e^{(r-\delta)(T-s)}+\beta e^{-\rho s}U(c^*(s))+2\delta b(s)+2\delta\beta q(s)\right]e^{-2\delta s}ds.\\
\end{aligned}
\end{equation}
Therefore, the solution to the mean-variance-utility problem can be  presented in the following proposition.
\begin{proposition}\label{prop_1}
\textcolor{blue}{Suppose that utility function $U(\cdot)$ is continuous, concave and increasing.} The optimal consumption and investment strategy for problem \eqref{maxu22} are respectively given by
\begin{equation}\label{o3a}
c^*=[U']^{-1}\left(\beta^{-1}e^{r(T-t)-\delta T}\right)
\end{equation}
and
\begin{equation}\label{o3b}
\pi^*=\frac{1}{\gamma}\frac{\mu-r}{x\sigma^2}e^{-(r-\delta)(T-t)}.
\end{equation}
In particular, when the discount rate $\delta$ is 0, we can obtain
\begin{equation}\label{o3aa}
c^*=[U']^{-1}\left(\beta^{-1}e^{r(T-t)}\right)
\end{equation}
and
\begin{equation}\label{o3bb}
\pi^*=\frac{1}{\gamma}\frac{\mu-r}{x\sigma^2}e^{-r(T-t)},
\end{equation}
which are exactly the desired optimal strategy of the mean-variance-utility problem \eqref{maxu2}.
This shows that under the mean-variance-utility criterion, the optimal consumption rate of the investor is independent of the current wealth, while the optimal investment rate is reversely proportional to the current wealth.
\end{proposition}

%It should be pointed out that the specific objective of the mean-variance-utility problem \eqref{maxu2} considered in this paper actually belongs to the category of time-inconsistent control problems, which means that it can also be solved with the general framework proposed by Bj\"{o}rk et al. \cite{Bjork17}. For the completeness of the paper and easiness of reference, the alternative derivation is also included in the Appendix. On the other hand, in addition to the mathematical theory in solving this specific problem, we also try to obtain some useful conclusions from the economical point of view. Our main emphasis is to formulate an interesting economical problem and attempt to provide the clear strategies for a new kind of investment problem, which hasn't been studied before.

\begin{remark}

\rm The new optimal portfolio selection problem subject to a minimized risk at the end of an investment period has led to at least two very interesting features that clearly distinguish themselves from those of the Merton's classic framework \cite{Merton75}:

i) The optimal consumption strategy derived under the current mean-variance-utility framework is independent of the wealth, as suggested by Eq. \eqref{o3aa}. In the Merton's classic framework, the optimal consumption depends on one's current total accumulated wealth. This of course makes sense economically as one would probably feel that he/she can afford to consume more when his/her total wealth is larger. However, in our new problem, the newly introduced risk control at the end of the investment period has magically balanced out such a dependence; our solution Eq. \eqref{o3aa} suggests that one's optimal consumption should be independent of the current wealth and be an increaseing function of $t$. But, his/her consumption is no longer directly proportional to the total wealth, as there is not much time left at the end of an investment period to control the total investment risk while optimizing his/her return. When there is no need to worry about the investment risk at all, his/her investment behavior would naturally be different as suggested by Merton's original framework. \textcolor{blue}{It should also be noted that the final wealth under our framework can be negative with positive probability. This is because we have not imposed additional constraints of a non-negative final wealth. As presented by \cite{Korn95}, for our objective utility function, a non-negative wealth can only be achieved if one explicitly forces this as a side constraint in the optimization problem. However, with this constraint, one will not obtain the explicit results and similar results can be found in \cite{Korn95,Korn10,Alp13}.}

ii)On the other hand, the optimal investment strategy in our problem, \eqref{o3bb}, shows that the optimal investment strategy $\pi^*$ obtained in this paper is dependent of the current wealth, which is consistent with the reality. The optimal investment rate is in fact inversely related to the current wealth value, since investors have to manage the risk of the current wealth under the mean-variance criterion, in which case the investment rate will be slowed down when the wealth value increases. If we further rewrite \eqref{o3bb} as
\begin{equation}\label{o3b-}
\pi^*x=\frac{1}{\gamma}\frac{\mu-r}{\sigma^2}e^{-r(T-t)}.
\end{equation}
\textcolor{blue}{it is not difficult to find that the dollar amount invested in the risky asset at time $t$ is independent of the current wealth $x$, which agrees well with previous relevant works in the literature \cite{Basak10,Bjork12,Kronborg15}}. The most astonishing part, however, is that all the optimal investment ends with a same form, as long as a mean variance is built into a model. Specifically, the optimal investment found by Basak and Chabakauri \cite{Basak10}, who solve the dynamic mean-variance portfolio problem and derive its time-consistent
solution using dynamic programming, and Bj\"{o}rk et al. \cite{Bjork12}, who placing the mean-variance problem within a game theoretic framework and obtain the subgame perfect Nash
equilibrium strategies with time inconsistency, and even
Kronborg and Steffensen \cite{Kronborg15}, who take into account the consumption term in mean-variance framework and consider the optimal investment strategy, all share a common formula, Eq. \eqref{o3bb}, no matter where the mean variance is placed at. \textcolor{blue}{This suggests that mean-variance term added to minimize the investment risk has only exerted influence on the investment strategy itself and has not strictly restricted the choice of consumption.} This makes economical sense as risks associated with investing in risky asset are a different type of risk from those associated with consumption, which has a direct impact on the total wealth available to be invested at each point of the investment horizon.
\end{remark}

It should be noted that as the consumption is deterministic, the investor can actually just adjust the initial capital by the present value of future (deterministic) income and future (deterministic) consumption and then invest the remaining capital according to Basak and Chabakauri \cite{Basak10}. Since the Basak and Chabakauri strategy (amount invested) is independent of wealth, the consumption/investment combination actually has a simple and reasonable interpretation, which is stated below.
\begin{remark}
\rm
The optimal investment strategy obtained under our mean-variance-utility framework is exactly the same as that derived in \cite{Christiansen13,Kronborg15}, with the optimal amount of money being independent of wealth\footnote{As pointed out in \cite{Kronborg15}, this seems to be economically unreasonable for a multi-period model, and a possible way to resolve this issue is to make the risk aversion be time- and wealth-dependent. We refer interested readers to \cite{Kronborg15} for more detailed discussion.}. A possible explanation is that intermediary consumption is independent of wealth and involves no risk, which indicates that the structure of the solution to the remaining mean-variance problem remains the same, independent of this consumption term. In particular, one has to firstly finance the deterministic optimal consumption, and the rest of the capital is invested according to a mean-variance problem without consumption. As the capital is invested independently of the size of the wealth, financing optimal consumption does not play a role there. Therefore, the presence of consumption utility could not yield the fundamental change in the optimal investment strategy, and the form of the investment strategy remains as the same as that under the mean-variance criterion. We also note that, the optimal consumption strategies under the two frameworks are completely different, as the one in \cite{Christiansen13,Kronborg15} is discrete, taking either the maximal or minimal allowed value, while ours is continuous. Moreover, from an economic point of view, their results are actually not reasonable as it is usually not possible for a normal investor to make sudden changes in his/her consumption strategy from consuming the maximal to the minimal allowed value.
\end{remark}

Having successfully derived the optimal consumption and investment strategy, it is not difficult to formulate the optimal value function of the problem \eqref{maxu22} as
\begin{equation}\label{valueaa}
V(t,x)=e^{(r-\delta)(T-t)}x+B(t),
\end{equation}
where $B(t)$ is specified in \eqref{BB}. Obviously, the optimal value function at the terminal point is constructed with the accumulated amount of the wealth $x$ at time $t$ and the additional amount resulted from the continuous consumption and investment strategy. With
\begin{equation}
V_x(t,x)=e^{(r-\delta)(T-t)}>0,
\end{equation}
the optimal value increases with wealth, which financially matches with one's intuition. The sensitivity of the optimal value function with respect to the time is affected by two aspects, i.e., the wealth, and the consumption and investment strategy.

The optimal strategy derived in \eqref{o3a} and \eqref{o3b} can also give rise to the conditional expected value and conditional second moment of the discounted optimal terminal wealth, yielding
\begin{equation}
E\left[\left.e^{-\delta(T-t)}X^{c^*,\pi^*}(T)\right|X(t)=x\right]=e^{(r-\delta)(T-t)}x+b(t),
\end{equation}
and
\begin{equation}
E\left[\left.\left(e^{-\delta(T-t)}X^{c^*,\pi^*}(T)\right)^2\right|X(t)=x\right]=\frac{2}{\gamma}\left\{b(t)-B(t)+\beta\left[p(t)x+q(t)\right]\right\},
\end{equation}
respectively. One can also similarly compute the conditional expectation of the discounted accumulated utility of consumption as
\begin{equation}
E\left[\left.\int_{t}^{T}e^{-\rho(T-t)}U(c^*(s))\right|X(t)=x\right]=\int_{t}^{T}e^{-\rho s}U(c^*(s))ds.
\end{equation}

To further investigate the properties of the optimal consumption strategy as well as the corresponding optimal value function, we now provide three examples with specific utility functions.
\begin{proposition}\label{prop_u}
With some particular choices of utility functions for problem \eqref{maxu}, the corresponding consumption strategies can be specified according to Proposition \ref{prop_1}.
\begin{itemize}
\item [(i)] With a logarithmic utility function $U(c)=\log(c)$, the optimal consumption strategy \eqref{o3a} can be simplified to take the form
\begin{equation}\label{loga}
c^*=\beta e^{-r(T-t)+\delta T}.
\end{equation}
Then,
\begin{equation}\label{logaa}
c^*=\beta e^{-r(T-t)}
\end{equation}
is the optimal one for the mean-variance-utility problem \eqref{maxu2}.
\item [(ii)] With a power utility function $U(c)=c^{\theta}/\theta$, where $\theta<1$ and $\theta\neq 0$, the optimal consumption strategy \eqref{o3a} becomes
\begin{equation}
c^*=\left(\beta^{-1}e^{r(T-t)-\delta T}\right)^{\frac{1}{\theta-1}}.
\end{equation}
Then,
\begin{equation}
c^*=\left(\beta^{-1}e^{r(T-t)}\right)^{\frac{1}{\theta-1}}
\end{equation}
is the optimal one for the mean-variance-utility problem \eqref{maxu2}.
\item [(iii)] With an exponential utility function $U(c)=-e^{-\eta c}/\eta$ with $\eta>0$, the optimal consumption strategy \eqref{o3a} can be explicitly obtained as
\begin{equation}
c^*=\frac{1}{\eta}\left[\ln \beta-r(T-t)+\delta T\right].
\end{equation}
Then,
\begin{equation}
c^*=\frac{1}{\eta}\left[\ln \beta-r(T-t)\right]
\end{equation}
is the optimal one for the mean-variance-utility problem \eqref{maxu2}.
\end{itemize}
\end{proposition}

With different optimal consumption strategies being derived corresponding to different utility functions, it is of interest to investigate the effect of the newly introduced parameter, $\beta$, on the optimal objective value function $V(t,x)$. Let's adopt the logarithmic utility function to mean-variance-utility problem \eqref{maxu2} (setting $\delta=0$ in \eqref{BB} and \eqref{valueaa}) as an example for illustration. Substituting \eqref{logaa} into \eqref{valueaa} yields
\begin{equation}
\frac{dV}{d\beta}=M(t)+M(t)\log\beta
\end{equation}
and
\begin{equation}
\frac{d^2V}{d\beta^2}=\frac{1}{\beta}M(t),
\end{equation}
where
$$M(t)=\int_{t}^{T}e^{-\rho s}[-r(T-s)+\rho T]ds.$$
From the expression of the first-order derivative, one can easily observe that the changes of the value function with respect to $\beta$ are dependent on both the parameter values and some other time-dependent functions, which implies that the sensitivity of the value function over $\beta$ will be adjusted over time. In addition, it is not difficult to find that $V_{\beta\beta}>0$ when $r<\rho$, which suggests that the sensitivity of the value function towards $\beta$ is a monotonic increasing function of $\beta$ in the case the expected return on the saved money is less than expected return on consumption utility. This is also reasonable, as $\beta$ denotes the consumption preference, and when $\beta$ is large, any tiny changes in $\beta$ value would result in a large change in the consumption strategy, leading to a significant impact on the value function.

Apart from the optimal value function, one may also be interested to see how the optimal strategies behave with respect to different parameter values, the details of which are provided in the next section. Before we finish here, it should be remarked that we may not always be able to derive the specific form of the optimal consumption strategy, since it depends on whether we can find the inverse of the derivative of the chosen utility function. However, even when the analytical inversion of the selected utility function is not available, e.g., when some mixed utility functions are adopted, it is still very straightforward to implement \eqref{o3a} in some numerical softwares like Matlab to compute.

One may wonder what happens if $\beta$ goes to infinity. Mathematically, this limit process will lead to an ill-posed problem, as far as the optimization is concerned. In fact, the infinite $\beta$ value will lead to abnormal (infinite) consumption, which can only be allowed if the income of investors is also abnormal (infinite). If we assume that investors only have limited initial wealth and normal income, then in order to maintain the balance of income and expenditure of investors, $\beta$ should be constrained to a reasonable but not infinite range. Financially, such a limit has actually freed the investor from the ``hassle" of trying to optimize his/her portfolio in the sense that he/she could consume without restraints which is actually not reasonable as the investor would normally keep a balanced budget. If $\beta$ goes to infinity, the optimal consumption will also approach infinity. This is because $\beta$ going to infinity means that the investor does not care about final wealth and the mean-variance concern for terminal wealth becomes redundant, in which case the investor will consume as much as possible.

\section{Numerical examples}

\textcolor{blue}{In this section, the properties of  optimal consumption strategy under three common utility functions discussed in Proposition 4.2 are investigated. Numerical examples and detailed discussions are provided by setting $\mu$ and $r$ as $0.05$ and $0.01$, respectively.}

The optimal strategy in this paper is applicable to general utility function, as long as the basic definition of utility function is satisfied: the more satisfied a person is with consumption, the better, that is, the first derivative of utility function is greater than zero; with \textcolor{blue}{the increased consumption, the speed of satisfaction decreases}, and the second derivative of utility function is less than zero. Once a specific utility function is selected, the optimal investment strategy will be determined. Now, we choose some simple and representative utility functions in economics to illustrate.

\begin{figure}[H]\center
\includegraphics[width=.54\textwidth]{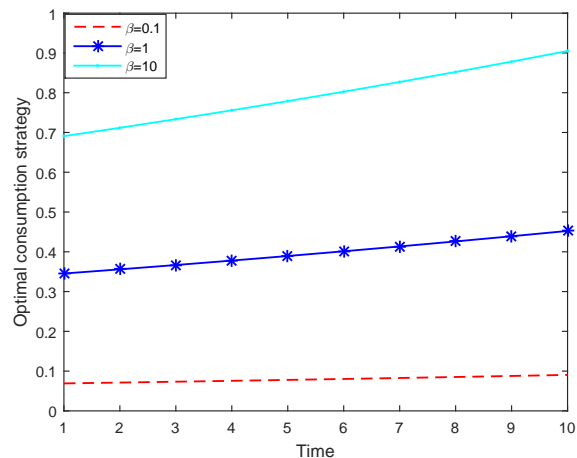}
\caption{Optimal consumption under a logarithmic utility function.}
\label{log_fig}
\end{figure}

First of all, depicted in Figure \ref{log_fig} is the optimal consumption strategy with different $\beta$ values when the investor chooses a logarithmic utility function. One can easily observe that the investor tends to consume more when the $\beta$ value is higher. This is indeed reasonable as an increase in $\beta$ places a higher weight on the accumulated utility when calculating the value function, and this corresponds to the case where the investor \textcolor{blue}{prefers to increase consumption} to achieve a higher utility than managing wealth under the mean-variance framework. It is also interesting to find that the investor would like to raise the level of consumption when the end of the pre-determined investment period is approached under our mean-variance-utility framework. This may appear to be strange at a first glance, but this could also be understood from an economic point of view. At the early stage, a rational investor tends to be conservative in terms of consumption given that there may be plenty of uncertainty with maximizing the terminal wealth being part of his/her long-term goal in achieving maximum ``happiness", and thus managing his/her terminal wealth through investment has higher priority over consumption. However, when the time passes by and the investor has accumulated certain amount of wealth, he/she would gain more confidence in consuming more to more ``happiness". This is indeed consistent with our theoretical findings for the optimal value function.

\begin{figure}[h]\center
\includegraphics[width=.48\textwidth]{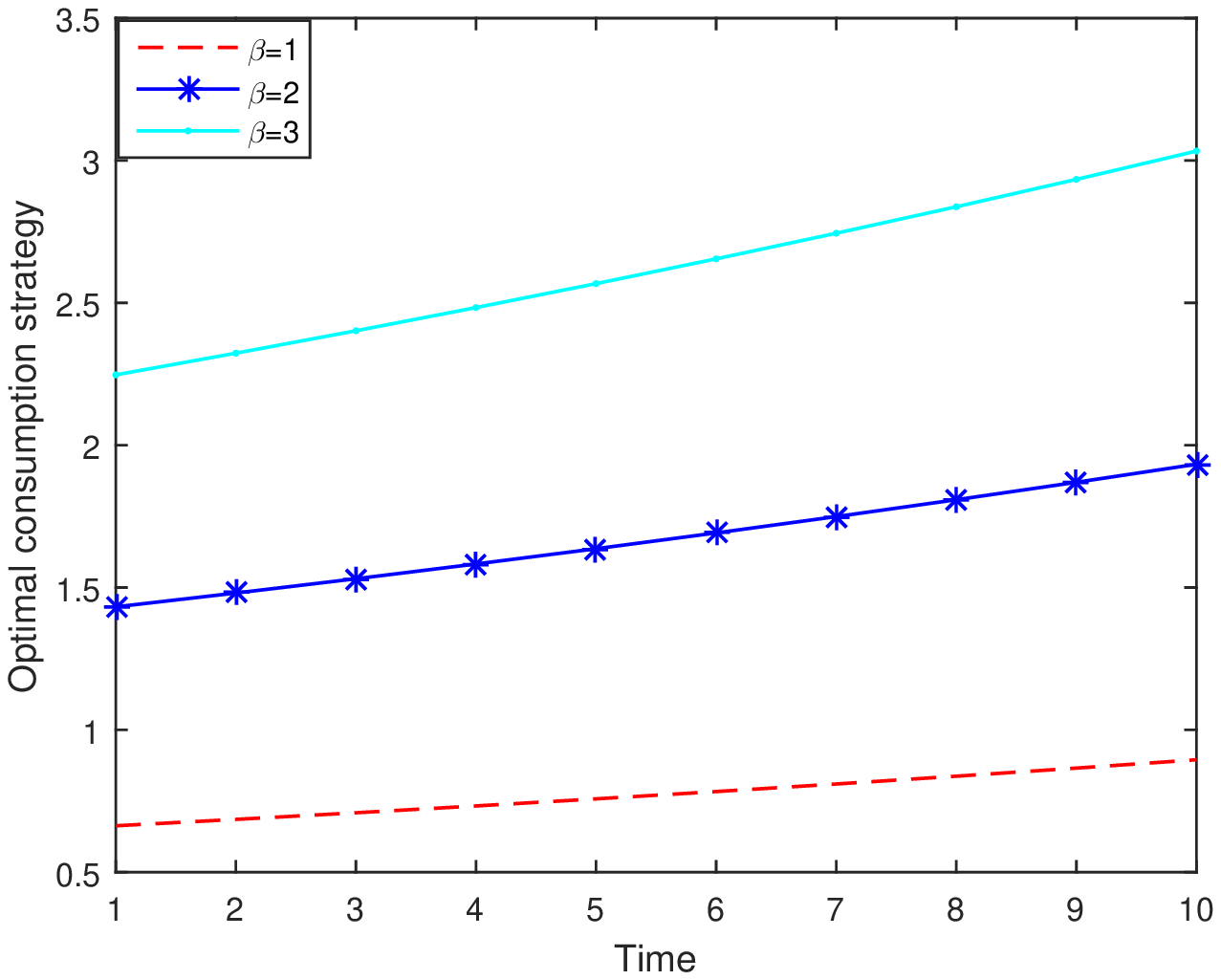}
\includegraphics[width=.48\textwidth]{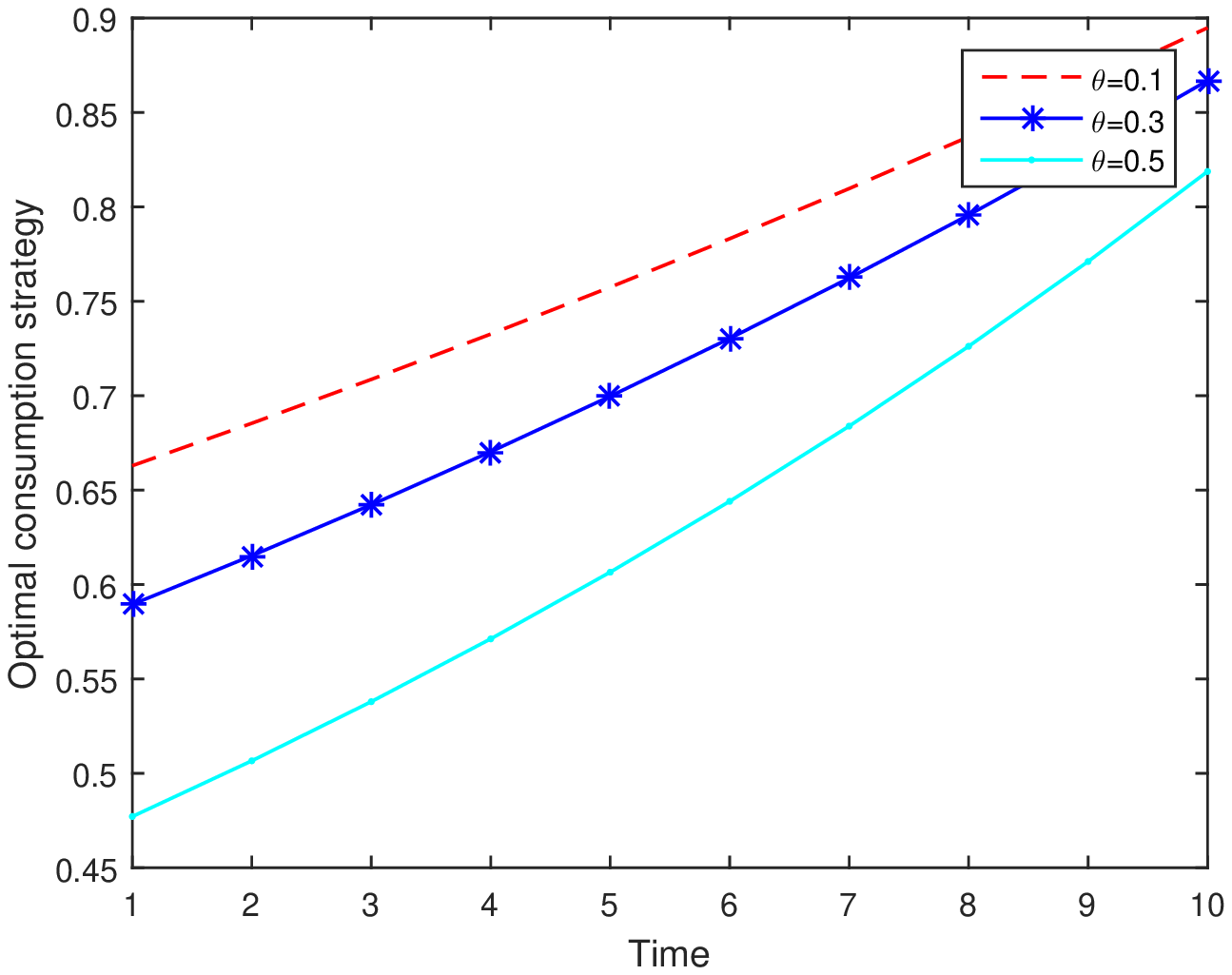}
\caption{Optimal consumption under the power utility function. $\theta$ used in left subfigure and $\beta$ used in the right subfigure are set to be 0.1 and 1, respectively.}
\label{power}
\end{figure}

\begin{figure}[h]\center
\includegraphics[width=.48\textwidth]{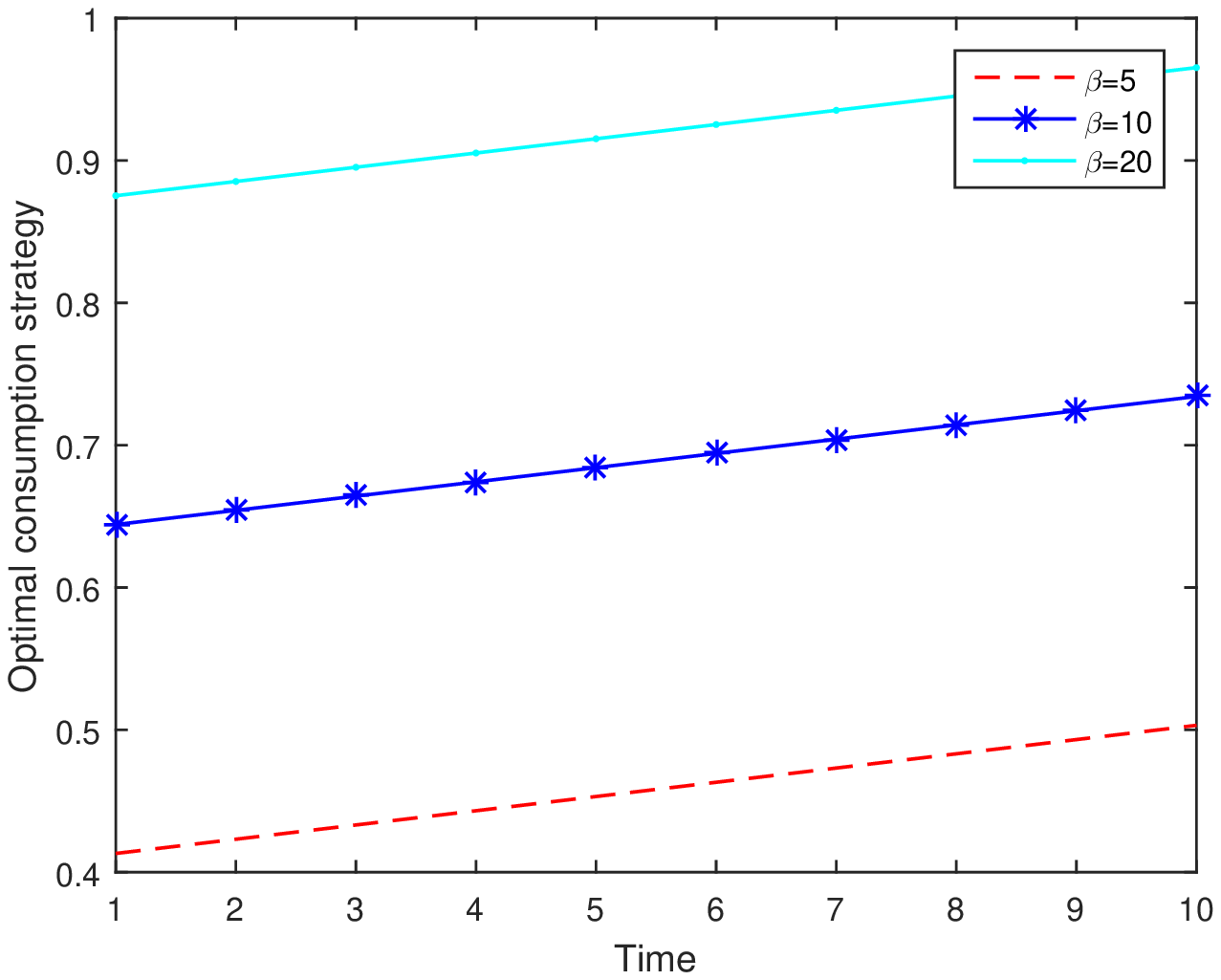}
\includegraphics[width=.48\textwidth]{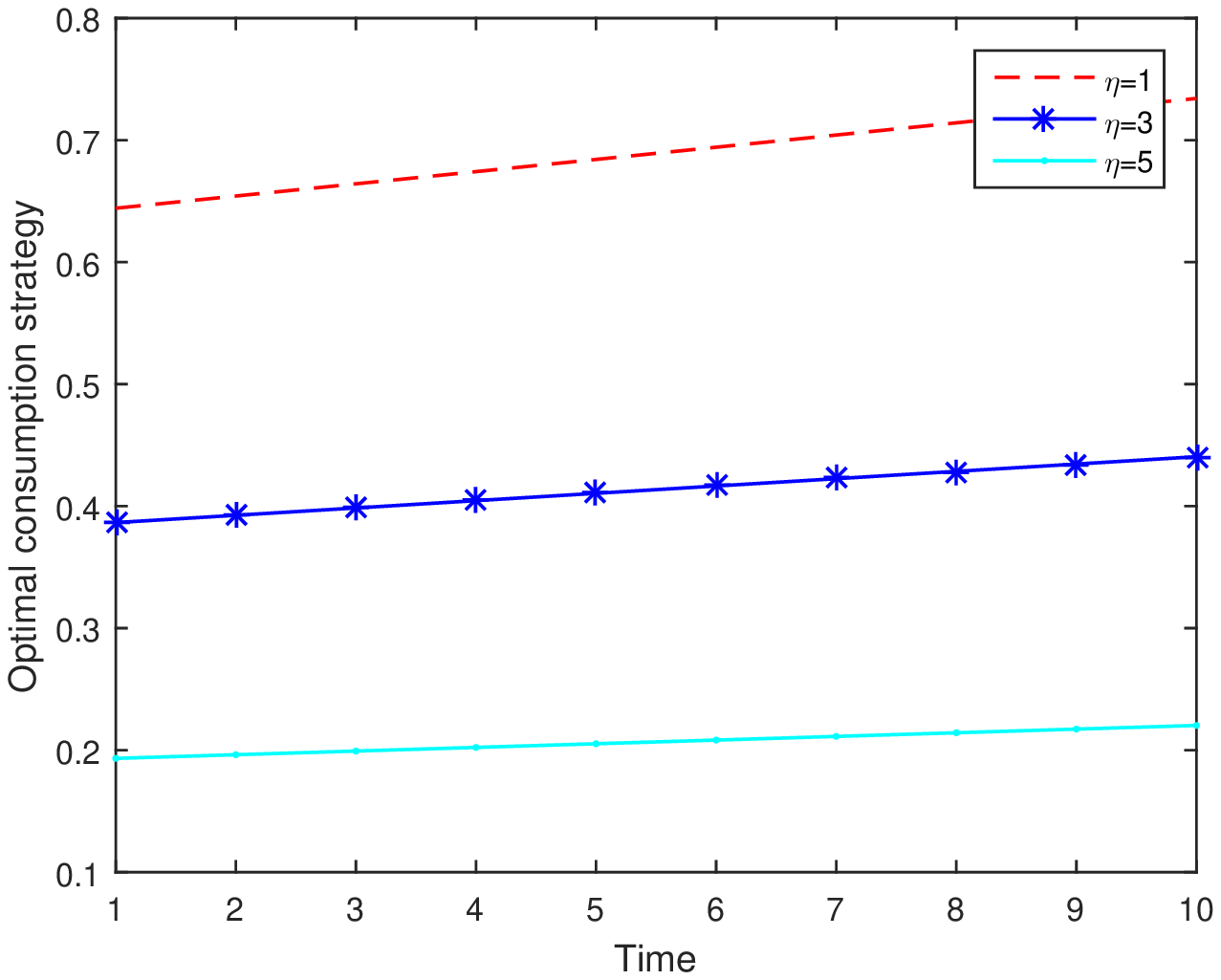}
\caption{Optimal consumption under the exponential utility function. \textcolor{blue}{The fixed parameter $\eta$ used in left subfigure and fixed parameter $\beta$ used in the right subfigure are set to be 3 and 10, respectively.}}
\label{exp}
\end{figure}

Figures \ref{power} and \ref{exp} display how the optimal consumption strategy varies when the utility function is in the form of a power and an exponential function, respectively. What can be observed first from both figures is a similar pattern as shown in Figure \ref{log_fig} that the optimal consumption strategy is still a monotonic increasing function of $\beta$, as a higher value of $\beta$ still implies that more consumption is preferred. Another phenomenon that should be noted is that the investor is willing to consume more when $\theta$ ($\eta$) takes smaller values. The main explanation for this is that $\theta$ ($\eta$) indicates the degree of risk aversion, and a larger $\theta$ ($\eta$) value implies avoiding excessive consumptions.

\begin{figure}[h]\center
\includegraphics[width=.48\textwidth]{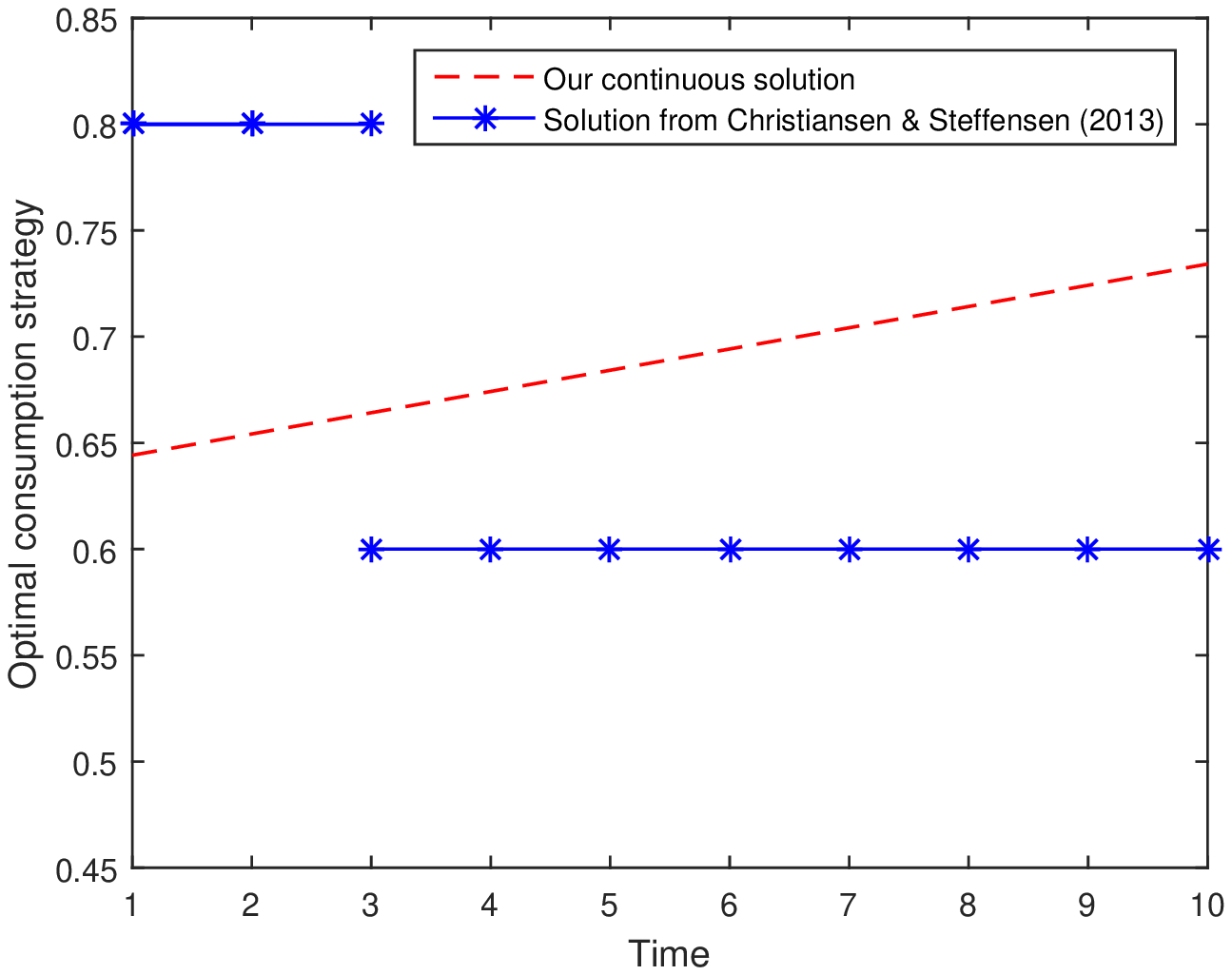}
\includegraphics[width=.48\textwidth]{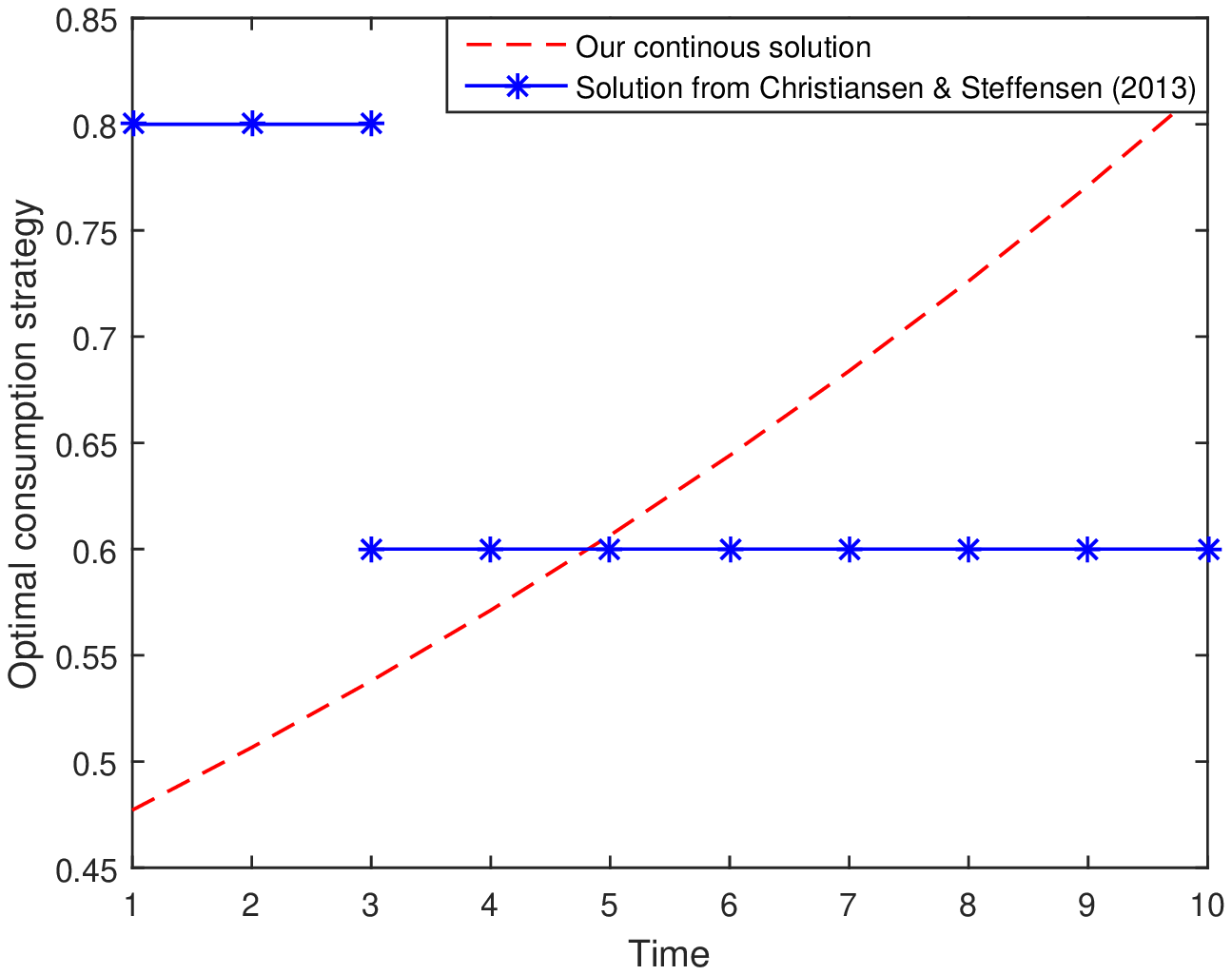}
\caption{\textcolor{blue}{Comparisons of the optimal consumption strategies.} }
\label{compare}
\end{figure}

It is also interesting to show the difference between the optimal consumption strategy derived under our framework and that obtained in \cite{ Christiansen13}, as there are two different approaches used to incorporate the consumption into the mean-variance problem. In particular, the problem proposed in \cite{Christiansen13} does not admit an explicit and analytical solution, and it was to be numerically solved with the fixed-point method, while our optimal consumption strategy is completely closed form solution, which facilitates  its practical applications. \textcolor{blue}{For comparison, we respectively choose the power utility function with $\theta=0.5$ and $\beta=1$ and the exponential utility function with $\eta=1$ and $\beta=10$ in our model. As displayed in Figure \ref{compare}, the optimal consumption strategy derived in \cite{Christiansen13} is not continuous, and there exists a sudden drop from the maximal to minimal allowed consumption rate for any investor using their framework.} This is by no means reasonable, since an investor would never consider to make substantial changes in his/her consumption in normal situations. \textcolor{blue}{As depicted in Figure \ref{compare}, our optimal consumption strategy, on the other hand, turns out to be continuous without break point}, being a monotonic increasing function of the time, which is more reasonable for the same reason stated above.

\section{Concluding remarks}

In this paper, we introduce the concept of overall ``happiness" of an investor, with which the terminal wealth under the mean-variance criterion and accumulated consumption utility can be directly added together using a consumption preference parameter, \textcolor{blue}{to formulate a new class of investment-consumption optimization problems as \eqref{maxu2}}. The optimal consumption strategy is continuous and increases over time, which is consistent with financial intuition that a normal investor would prefer investment over consumption when it is far away from the end of period to achieve more happiness, while he/she would gradually increase the level of consumption when the wealth starts to be accumulated.

\section*{\textcolor{blue}{Appendix}}

\subsection*{Appendix A: The proof of Lemma \ref{lemma2}}
\begin{proof}
Define
\begin{equation}\label{kkkk}
\widetilde{Y}(t,x)=e^{-\delta t}Y(t,x).
\end{equation}
\textcolor{blue}{Combining \eqref{kkkk} with \eqref{Y1} and then taking derivative, we can yield}
\begin{equation}\label{Y2}
\left\{
\begin{aligned}
\widetilde{Y}_t(t,x)&=-[(r+\pi(\mu-r))x+l-c]\widetilde{Y}_x(t,x)-\frac{1}{2}\pi^2\sigma^2x^2\widetilde{Y}_{xx}(t,x),\\
\widetilde{Y}(T,x)&=e^{-\delta T}x.
\end{aligned}
\right.
\end{equation}
Applying It\^o's lemma further yields
\begin{equation}\label{Y2}
\begin{aligned}
\widetilde{Y}(t,X^{c,\pi}(t))=&-\int_{t}^{T}d\widetilde{Y}(s,X^{c,\pi}(s))+\widetilde{Y}(T,X^{c,\pi}(T))\\
=&-\int_{t}^{T}\bigg(\widetilde{Y}_s(s,X^{c,\pi}(s))+[(r+\pi(s)(\mu-r))X^{c,\pi}(s)+l(s)-c(s)]\widetilde{Y}_x(s,X^{c,\pi}(s))\\
  &+\frac{1}{2}\pi^2(s)\sigma^2(X^{c,\pi}(s))^2\widetilde{Y}_{xx}(s,X^{c,\pi}(s))\bigg)ds
  -\int_{t}^{T}\pi(s)\sigma X^{c,\pi}(s)\widetilde{Y}_{x}(s,X^{c,\pi}(s))dB(s)\\
  &+\widetilde{Y}(T,X^{c,\pi}(T))\\
=&e^{-\delta T}X^{c,\pi}(T)-\int_{t}^{T}\pi(s)\sigma X^{c,\pi}(s)\widetilde{Y}_{x}(s,X^{c,\pi}(s))dB(s).
\end{aligned}
\end{equation}
Since $(c,\pi)$ is an admissible strategy, taking the expectation on two sides of \eqref{Y2} conditional upon $X(t)=x$ results in
\begin{equation}
\widetilde{Y}(t,X^{c,\pi}(t))=E\left[\left.e^{-\delta T}X^{c,\pi}(T)\right|X(t)=x\right],
\end{equation}
from which one can  obtain
\begin{equation}
Y(t,x)=e^{\delta t}\widetilde{Y}(t,x)=y^{c,\pi}(t,x).
\end{equation}
Similarly, if we denote
\begin{equation}
\widetilde{Z}(t,x)=e^{-2\delta t}Z(t,x),
\end{equation}
we can obtain
\begin{equation}\label{Z2}
\left\{
\begin{aligned}
\widetilde{Z}_t(t,x)&=-[(r+\pi(\mu-r))x+l-c]\widetilde{Z}_x(t,x)-\frac{1}{2}\pi^2\sigma^2x^2\widetilde{Z}_{xx}(t,x),\\
\widetilde{Z}(T,x)&=e^{-2\delta T}x^2.
\end{aligned}
\right.
\end{equation}
Again, applying It\^o's lemma leads to
\begin{equation}\label{Z3}
\begin{aligned}
\widetilde{Z}(t,X^{c,\pi}(t))=&-\int_{t}^{T}d\widetilde{Z}(s,X^{c,\pi}(s))+\widetilde{Z}(T,X^{c,\pi}(T))\\
=&-\int_{t}^{T}\bigg(\widetilde{Z}_s(s,X^{c,\pi}(s))+[(r+\pi(s)(\mu-r))X^{c,\pi}(s)+l(s)-c(s)]\widetilde{Z}_x(s,X^{c,\pi}(s))\\
  &+\frac{1}{2}\pi^2(s)\sigma^2(X^{c,\pi}(s))^2\widetilde{Z}_{xx}(s,X^{c,\pi}(s))\bigg)ds
  -\int_{t}^{T}\pi(s)\sigma X^{c,\pi}(s)\widetilde{Z}_{x}(s,X^{c,\pi}(s))dB(s)\\
  &+\widetilde{Z}(T,X^{c,\pi}(T))\\
=&\left(e^{-\delta T}X^{c,\pi}(T)\right)^2-\int_{t}^{T}\pi(s)\sigma X^{c,\pi}(s)\widetilde{Z}_{x}(s,X^{c,\pi}(s))dB(s).
\end{aligned}
\end{equation}
Taking the expectation on both sides of \eqref{Z3} conditional upon $X(t)=x$, it is straightforward that
\begin{equation}
\widetilde{Z}(t,X^{c,\pi}(t))=E\left[\left.\left(e^{-\delta T}X^{c,\pi}(T)\right)^2\right|X(t)=x\right],
\end{equation}
and thus
\begin{equation}
Z(t,x)=e^{2\delta t}\widetilde{Z}(t,x)=z^{c,\pi}(t,x).
\end{equation}
Finally, following a similar fashion, $W(t,X^{c,\pi}(t))$ satisfying \eqref{W1} can be founded as
\begin{equation}\label{W3}
\begin{aligned}
W(t,X^{c,\pi}(t))=&-\int_{t}^{T}dW(s,X^{c,\pi}(s))+W(T,X^{c,\pi}(T))\\
=&-\int_{t}^{T}\bigg(W_s(s,X^{c,\pi}(s))+[(r+\pi(s)(\mu-r))X^{c,\pi}(s)+l(s)-c(s)]W_x(s,X^{c,\pi}(s))\\
  &+\frac{1}{2}\pi^2(s)\sigma^2(X^{c,\pi}(s))^2W_{xx}(s,X^{c,\pi}(s))\bigg)ds
  -\int_{t}^{T}\pi(s)\sigma X^{c,\pi}(s) W_{x}(s,X^{c,\pi}(s))dB(s)\\
  &+W(T,X^{c,\pi}(T))\\
=&\int_{t}^{T}e^{-\rho s}U(c(s))ds-\int_{t}^{T}\pi(s)\sigma X^{c,\pi}(s) \widetilde{W}_{x}(s,X^{c,\pi}(s))dB(s).
\end{aligned}
\end{equation}
Taking the conditional expectation on (\ref{W3}) yields the desired result. This has completed the proof.
\end{proof}

\subsection*{Appendix B: The proof of Lemma \ref{lemma2}}
\begin{proof}
The proof process is divided into three steps. The first step is to derive an expression for
\begin{equation}\label{f1}
f^{c,\pi}(t,X^{c,\pi}(t),y^{c,\pi}(t,X^{c,\pi}(t)),z^{c,\pi}(t,X^{c,\pi}(t)),w^{c,\pi}(t,X^{c,\pi}(t))).
\end{equation}
Using It\^o's lemma, we have
\begin{equation}\label{0f2}
\begin{aligned}
& f^{c,\pi}(t,X^{c,\pi}(t),y^{c,\pi}(t,X^{c,\pi}(t)),z^{c,\pi}(t,X^{c,\pi}(t)),w^{c,\pi}(t,X^{c,\pi}(t))\\
\quad=&-\int_{t}^{T}df^{c,\pi}(t,X^{c,\pi}(t),y^{c,\pi}(t,X^{c,\pi}(t)),z^{c,\pi}(t,X^{c,\pi}(t)),w^{c,\pi}(t,X^{c,\pi}(t))\\
&+f^{c,\pi}(T,X^{c,\pi}(T),y^{c,\pi}(T,X^{c,\pi}(T)),z^{c,\pi}(T,X^{c,\pi}(T)),w^{c,\pi}(t,X^{c,\pi}(T))\\
\quad=&-\int_{t}^{T}\bigg\{(f_s^{c,\pi}+f_y^{c,\pi}Y_s+f_z^{c,\pi}Z_s+f_w^{c,\pi}W_s)ds+(f_x^{c,\pi}+f_y^{c,\pi}Y_x+f_z^{c,\pi}Z_x++f_w^{c,\pi}W_x)dX^{c,\pi}(s)\\
&+\frac{1}{2}\pi^2(s)\sigma^2(X^{c,\pi}(s))^2\big[f_y^{c,\pi}Y_{xx}+f_z^{c,\pi}Z_{xx}+f_w^{c,\pi}W_{xx}+f_{yy}^{c,\pi}(Y_{x})^2+
+f_{zz}^{c,\pi}(Z_{x})^2+f_{ww}^{c,\pi}(W_{x})^2\\
&+2f_{xy}^{c,\pi}Y_x+2f_{xz}^{c,\pi}Z_x+2f_{xw}^{c,\pi}W_x+2f_{yz}^{c,\pi}Y_xZ_x+2f_{yw}^{c,\pi}Y_xW_x+2f_{zw}^{c,\pi}Z_xW_x\big]\bigg\}ds\\
&+f^{c,\pi}(T,X^{c,\pi}(T),Y^{c,\pi}(T,X^{c,\pi}(T)),Z^{c,\pi}(T,X^{c,\pi}(T)),W^{c,\pi}(T,X^{c,\pi}(T)).
\end{aligned}
\end{equation}
Using \eqref{Y1}, \eqref{Z1} and \eqref{W1}, we further have
\begin{equation}\label{1f2}
\begin{aligned}
&f^{c,\pi}(t,X^{c,\pi}(t),y^{c,\pi}(t,X^{c,\pi}(t)),z^{c,\pi}(t,X^{c,\pi}(t)),w^{c,\pi}(t,X^{c,\pi}(t))\\
\quad=&-\int_{t}^{T}\bigg\{f_s^{c,\pi}ds+f_y^{c,\pi}\left(-[(r+\pi(\mu-r))X^{c,\pi}(s))+l-c]Y_x-\frac{1}{2}\pi^2\sigma^2(X^{c,\pi}(s))^2Y_{xx}+\delta Y\right)\\
&+f_z^{c,\pi}\left(-[(r+\pi(\mu-r))X^{c,\pi}(s))+l-c]Z_x-\frac{1}{2}\pi^2\sigma^2(X^{c,\pi}(s))^2Z_{xx}+2\delta Z\right)\\
&+f_w^{c,\pi}\left(-[(r+\pi(\mu-r))X^{c,\pi}(s))+l-c]W_x-\frac{1}{2}\pi^2\sigma^2(X^{c,\pi}(s))^2W_{xx}-e^{-\rho s}U(c(s)\right)\\
&+(f_x^{c,\pi}+f_y^{c,\pi}Y_x+f_z^{c,\pi}Z_x+f_w^{c,\pi}W_x)\left([(r+\pi(s)(\mu-r))X^{c,\pi}(s)+l(s)-c(s)]ds+\pi(s)\sigma X^{c,\pi}(s)dB(s)\right)\\
&+\frac{1}{2}\pi^2(s)\sigma^2(X^{c,\pi}(s))^2\big[f_y^{c,\pi}Y_{xx}+f_z^{c,\pi}Z_{xx}+f_w^{c,\pi}W_{xx}+f_{yy}^{c,\pi}(Y_{x})^2
+f_{zz}^{c,\pi}(Z_{x})^2+f_{ww}^{c,\pi}(W_{x})^2\\
&+2f_{xy}^{c,\pi}Y_x+2f_{xz}^{c,\pi}Z_x+2f_{xw}^{c,\pi}W_x+2f_{yz}^{c,\pi}Y_xZ_x+2f_{yw}^{c,\pi}Y_xW_x+2f_{zw}^{c,\pi}Z_xW_x\big]\bigg\}ds\\
&+f^{c,\pi}(T,X^{c,\pi}(T),Y^{c,\pi}(T,X^{c,\pi}(T)),Z^{c,\pi}(T,X^{c,\pi}(T)),W^{c,\pi}(T,X^{c,\pi}(T)).
\end{aligned}
\end{equation}
Therefore,
\begin{equation}\label{f2}
\begin{aligned}
&f^{c,\pi}(t,X^{c,\pi}(t),y^{c,\pi}(t,X^{c,\pi}(t)),z^{c,\pi}(t,X^{c,\pi}(t)),w^{c,\pi}(t,X^{c,\pi}(t))\\
\quad=&-\int_{t}^{T}\bigg\{(f_s^{c,\pi}+f_y^{c,\pi}\delta Y+2f_z^{c,\pi}\delta Z-f_w^{c,\pi}e^{-\rho s}K(c(s))ds\\
&+f_x^{c,\pi}[(r+\pi(t)(\mu-r))X^{c,\pi}(s)+l(s)-c(s)]ds\\
&+\pi(s)\sigma X^{c,\pi}(s)(f_x^{c,\pi}+f_y^{c,\pi}Y_x+f_z^{c,\pi}Z_x+f_w^{c,\pi}W_x)dB(s))\\
&+\frac{1}{2}\pi^2(s)\sigma^2(X^{c,\pi}(s))^2\big[f_{xx}^{c,\pi}+f_{yy}^{c,\pi}(Y_{x})^2
+f_{zz}^{c,\pi}(Z_{x})^2+f_{ww}^{c,\pi}(W_{x})^2\\
&+2f_{xy}^{c,\pi}Y_x+2f_{xz}^{c,\pi}Z_x+2f_{xw}^{c,\pi}W_x+2f_{yz}^{c,\pi}Y_xZ_x+2f_{yw}^{c,\pi}Y_xW_x+2f_{zw}^{c,\pi}Z_xW_x\big]\bigg\}ds\\
&+f^{c,\pi}(T,X^{c,\pi}(T),Y^{c,\pi}(T,X^{c,\pi}(T)),Z^{c,\pi}(T,X^{c,\pi}(T)),W^{c,\pi}(T,X^{c,\pi}(T)).
\end{aligned}
\end{equation}
For an arbitrary admissible strategy $(c,\pi)$, we furthermore define
\begin{equation}
\begin{aligned}
\widetilde{K}=&f_{xx}^{c,\pi}+f_{yy}^{c,\pi}(Y_{x})^2+f_{zz}^{c,\pi}(Z_{x})^2+f_{ww}^{c,\pi}(W_{x})^2+2f_{xy}^{c,\pi}Y_x+2f_{xz}^{c,\pi}Z_x\\
&+2f_{xw}^{c,\pi}W_x+2f_{yz}^{c,\pi}Y_xZ_x+2f_{yw}^{c,\pi}Y_xW_x+2f_{zw}^{c,\pi}Z_xW_x
\end{aligned}
\end{equation}
and
\begin{equation}
\widetilde{J}=f_t^{c,\pi}+f_y^{c,\pi}\delta Y+2f_z^{c,\pi}\delta Z-f_w^{c,\pi}e^{-\rho t}U(c(t)).
\end{equation}
This leads to
\begin{equation}\label{f2}
\begin{aligned}
& f^{c,\pi}(t,X^{c,\pi}(t),y^{c,\pi}(t,X^{c,\pi}(t)),z^{c,\pi}(t,X^{c,\pi}(t)),w^{c,\pi}(t,X^{c,\pi}(t))\\
\quad=&-\int_{t}^{T}\bigg\{(\widetilde{J}(s)+f_x^{c,\pi}[(r+\pi(t)(\mu-r))X^{c,\pi}(s)+l(s)-c(s)]ds\\
&+\pi(s)\sigma X^{c,\pi}(s)(f_x^{c,\pi}+f_y^{c,\pi}Y_x+f_z^{c,\pi}Z_x+f_w^{c,\pi}W_x)dB(s))\\
&+\frac{1}{2}\pi^2(s)\sigma^2(X^{c,\pi}(s))^2\widetilde{K}(s)ds\\
&+f^{c,\pi}(T,X^{c,\pi}(T),Y^{c,\pi}(T,X^{c,\pi}(T)),Z^{c,\pi}(T,X^{c,\pi}(T)),W^{c,\pi}(T,X^{c,\pi}(T)).
\end{aligned}
\end{equation}
With the utilization of It\^o's lemma, one can easily derive
\begin{equation}\label{F5}
\begin{aligned}
F(t,X^{c,\pi}(t))&=-\int_{t}^{T}dF(s,X^{c,\pi}(s))+F(T,X^{c,\pi}(T))\\
&=-\int_{t}^{T}\left(F_sds+F_xdX^{c,\pi}(s)+\frac{1}{2}F_{xx}(\pi(s))^2\sigma^2(X^{c,\pi}(s))^2ds\right)+F(T,X^{c,\pi}(T))
\end{aligned}
\end{equation}
Since $F$ solves the pseudo HJB equation \eqref{F1}, we can obtain that for any arbitrary strategy $(c,\pi)$, we have
$$F_t\leq-[(r+\pi(s)(\mu-r))x+l-c](F_x-Q)]-\frac{1}{2}\pi^2\sigma^2x^2(F_{xx}-K)+J.$$
Setting $x=X^{c,\pi}(s)$ in \eqref{wealth} and using the terminal conditions \eqref{Y1}, \eqref{Z1} and \eqref{W1} directly lead to
\begin{equation}\label{F5}
\begin{aligned}
F(t,X^{c,\pi}(t))\geq &-\int_{t}^{T}\bigg\{\big([(r+\pi(s)(\mu-r))X^{c,\pi}(s)+l(s)-c(s)](F_x-Q)\\
&-\frac{1}{2}\pi^2(s)\sigma^2(X^{c,\pi}(s))^2(-K(s)+F_{xx})+J(s)\big)ds\\
&+F_x\left([(r+\pi(s)(\mu-r))X^{c,\pi}(s)+l(s)-c(s)]ds+\pi(s)\sigma X^{c,\pi}(s)dB(s)\right)\\
&+\frac{1}{2}F_{xx}(\pi(s))^2\sigma^2(X^{c,\pi}(s))^2 ds\bigg\}\\
&+f^{c,\pi}(T,X^{c,\pi}(T),Y^{c,\pi}(T,X^{c,\pi}(T)),Z^{c,\pi}(T,X^{c,\pi}(T)),W^{c,\pi}(T,X^{c,\pi}(T))\\
= &-\int_{t}^{T}\bigg\{\big([(r+\pi(s)(\mu-r))X^{c,\pi}(s)+l(s)-c(s)]f_x^{c^*,\pi^*}(s)\\
&+\frac{1}{2}\pi^2(s)\sigma^2(X^{c,\pi}(s))^2K(s)+J(s)\big)ds+F_x\pi(s)\sigma X^{c,\pi}(s)dB(s)\bigg\}\\
&+f^{c,\pi}(T,X^{c,\pi}(T),Y^{c,\pi}(T,X^{c,\pi}(T)),Z^{c,\pi}(T,X^{c,\pi}(T)),W^{c,\pi}(T,X^{c,\pi}(T))\\
= &-\int_{t}^{T}\bigg\{\big([(r+\pi(s)(\mu-r))X^{c,\pi}(s)+l(s)-c(s)]\left(f_x^{c^*,\pi^*}(s)-f_x^{c,\pi}(s)\right)\\
&+\frac{1}{2}\pi^2(s)\sigma^2(X^{c,\pi}(s))^2(K(s)-\widetilde{K}(s))+J(s)-\widetilde{J}(s)\big)ds\\
&+\left(f_x^{c,\pi}+f_y^{c,\pi}Y_x+f_z^{c,\pi}Z_x+f_w^{c,\pi}W_x-F_x\right)\pi(s)\sigma X^{c,\pi}(s)dB(s)\bigg\}\\
&+f^{c,\pi}(t,X^{c,\pi}(t),y^{c,\pi}(t,X^{c,\pi}(t)),z^{c,\pi}(t,X^{c,\pi}(t)),w^{c,\pi}(t,X^{c,\pi}(t)),\\
\end{aligned}
\end{equation}
where the second equality follows from \eqref{f2}, after taking the expectation of the first equality conditional upon $X(t)=x$. We can thus arrive at
\begin{equation}\label{F4}
\begin{aligned}
f^{c,\pi}(t,x,y^{c,\pi}(t,x),z^{c,\pi}(t,x),w^{c,\pi}(t,x))\leq & F(t,x)+\int_{t}^{T}\bigg\{[(r+\pi(t)(\mu-r))X^{c,\pi}(s)+l(s)-c(s)]\\
& \times(f_x^{c^*,\pi^*}(s)-f_x^{c,\pi}(s))+J(s)-\widetilde{J}(s)\\
&+\frac{1}{2}\pi^2(s)\sigma^2(X^{c,\pi}(s))^2(K(s)-\widetilde{K}(s))\bigg\}ds.
\end{aligned}
\end{equation}

The last step is to check whether the Nash equilibrium criteria specified in Definition \ref{def1} are satisfied. If we assume that the strategy $(c^*,\pi^*)$ satisfies the infimum in \eqref{F1}, it follows from \eqref{s1} that
\begin{equation}\label{s2}
F^{(1)}(t,x)=y^{c^*,\pi^*}(t,x),\quad F^{(2)}(t,x)=z^{c^*,\pi^*}(t,x),\quad F^{(3)}(t,x)=w^{c^*,\pi^*}(t,x).
\end{equation}
As (\ref{F5}) holds for any admissible strategy $(c,\pi)$, it also applies for the specific strategy $(c^*,\pi^*)$, i.e.,
$$F_t=-[(r+\pi(\mu-r))x+l-c](F_x-Q)-\frac{1}{2}\pi^2\sigma^2x^2(F_{xx}-K)+J,$$
leading to
\begin{equation}\label{F6}
\begin{aligned}
F(t,X^{c^*,\pi^*}(t))= &\int_{t}^{T}\bigg\{\left(f_x^{c^*,\pi^*}+f_y^{c^*,\pi^*}Y_x+f_z^{c^*,\pi^*}Z_x+f_w^{c^*,\pi^*}W_x-F_x\right)\pi^*(s)\sigma X^{c^*,\pi^*}(s)dB(s)\bigg\}\\
&+f^{c^*,\pi^*}(t,X^{c^*,\pi^*}(t),y^{c^*,\pi^*}(t,X^{c^*,\pi^*}(t)),z^{c^*,\pi^*}(t,X^{c^*,\pi^*}(t)),w^{c^*,\pi^*}(t,X^{c^*,\pi^*}(t)).\\
\end{aligned}
\end{equation}
Taking the expectation on both sides of the above equality conditional upon $X(t)=x$ yields
\begin{equation}
F(t,x)=f^{c^*,\pi^*}\left(t,x,y^{c^*,\pi^*}(t,x),z^{c^*,\pi^*}(t,x),w^{c^*,\pi^*}(t,x)\right).
\end{equation}
If we consider the strategy $(\widetilde{c}_h,\widetilde{\pi}_h)$ defined in \eqref{pi0}, Equations \eqref{F4} and \eqref{F6} yield
\begin{equation*}
\begin{aligned}
&\lim_{h\rightarrow 0}\inf \frac{f^{c^*,\pi^*}(t,x,y^{c^*,\pi^*}(t,x),z^{c^*,\pi^*}(t,x),w^{c^*,\pi^*}(t,x))
-f^{\widetilde{c}_h,\widetilde{\pi}_h}(t,x,y^{\widetilde{c}_h,\widetilde{\pi}_h}(t,x),z^{\widetilde{c}_h,\widetilde{\pi}_h}(t,x),w^{\widetilde{c}_h,\widetilde{\pi}_h}(t,x))}{h}\\
&\geq  \lim_{h\rightarrow 0}\inf \frac{1}{h}\left\{\int_{t}^{T}\left[(r+\widetilde{\pi}_h(s)(\mu-r))X^{\widetilde{c}_h,\widetilde{\pi}_h}(s)+l(s)
-\widetilde{c}_h(s)\right](f_x^{\widetilde{c}_h,\widetilde{\pi}_h}(s)-f_x^{c^*,\pi^*}(s))ds\right.\\
&\quad\quad\left.+\int_{t}^{T}\left(\widetilde{J}_h(s)-J(s)+\frac{1}{2}\sigma^2(\widetilde{\pi}_h(s))^2(X^{\widetilde{c}_h,\widetilde{\pi}_h}(s))^2(\widetilde{K}_h(s)-K(s))ds\right)\right\}\\
&=  \lim_{h\rightarrow 0}\inf \frac{1}{h}\left\{\int_{t}^{t+h}\left[(r+\pi(s)(\mu-r))X^{c,\pi}(s)+l(s)
-c(s)\right](f_x^{\widetilde{c}_h,\widetilde{\pi}_h}(s)-f_x^{c^*,\pi^*}(s))ds\right.\\
& \quad\quad\left.+\int_{t}^{t+h}\left(\widetilde{J}_h(s)-J(s)+\frac{1}{2}\sigma^2(\pi(s))^2(X^{c,\pi}(s))^2(\widetilde{K}_h(s)-K(s))ds\right)\right\}\\
&=\left[(r+\pi(t)(\mu-r))X^{c,\pi}(t)+l(t)
-c(t)\right](f_x^{\widetilde{c}_0,\widetilde{\pi}_0}(t)-f_x^{c^*,\pi^*}(t))+ \widetilde{J}_0(t)-J(t)\\
&\quad+\frac{1}{2}\sigma^2(\pi(t))^2(X^{c,\pi}(t))^2(\widetilde{K}_0(t)-K(t))\\
&=0,
\end{aligned}
\end{equation*}
which implies that $F(t,x)=V(t,x)$ and $(c^*,\pi^*)$ is the desired optimal strategy.
\end{proof}
\section*{Acknowledgements}
\textcolor{blue}{The authors would like to gratefully acknowledge three anonymous referees' very valuable, technical and detailed comments and suggestions, which greatly help to improve the quality of the manuscript.}

\section*{Codes}
The codes for the numerical experiment are now provided below. All files are implemented by Matlab.

\begin{lstlisting}
function c=portfolio1(r,rho,mu)
T=[1:1:10];
for t=1:length(T);
A(t)=exp((r-rho)*(length(T)-t));
c(t)=1/(exp(rho*t)*A(t)*mu^(-1));% power utility function
end
\end{lstlisting}

\begin{lstlisting}
function c=portfolio2(r,rho,mu,theta)  % theta>0 <1
T=[1:1:10];
for t=1:length(T);
A(t)=exp((r-rho)*(length(T)-t));
c(t)=(exp(rho*t)*A(t)/mu)^(1/(theta-1)); % power utility function
end
\end{lstlisting}

\begin{lstlisting}
function c=portfolio3(r,rho,mu,eta)
T=[1:1:10];
for t=1:length(T);
A(t)=exp((r-rho)*(length(T)-t));
c(t)=-1/eta*log(exp(rho*t)*A(t)*mu^(-1));% exponential utility function
end
\end{lstlisting}

\end{document}